\documentclass[a4paper,11pt]{article}
\usepackage{pos}
\usepackage{subcaption}

\title{Polarised Beams at Future e$^+$e$^-$ Colliders}

\author*[a,1]{Jenny List}

\affiliation[a]{DESY, Notkestr. 85, 22607 Hamburg, Germany}

\note{{\bfseries \rm on behalf of the Linear Collider Collaboration }}

\emailAdd{jenny.list@desy.de}

\abstract{Beam polarisation is an integral part of the physics case of future Linear Colliders. In this contribution, important examples from Higgs coupling measurements, top and electroweak physics at high energies, the $Z$ pole program as well as from searches for production of new particles will be reviewed. The full exploitation of its advantages requires the polarisation to be known at the permille-level.
The polarimetry concept which has been developed for the ILC to
achieve the required precision based on the combination of Compton polarimeters, spin-tracking simulations and a global analysis of collision data will be presented.}

\FullConference{%
  40th International Conference on High Energy physics - ICHEP2020\\
  July 28 - August 6, 2020\\
  Prague, Czech Republic (virtual meeting)
}

\begin{document}
\maketitle

\section{Introduction}
\begin{figure}[htb]  
  \begin{subfigure}[h]{0.25\linewidth}
     \includegraphics[width=0.95\linewidth]{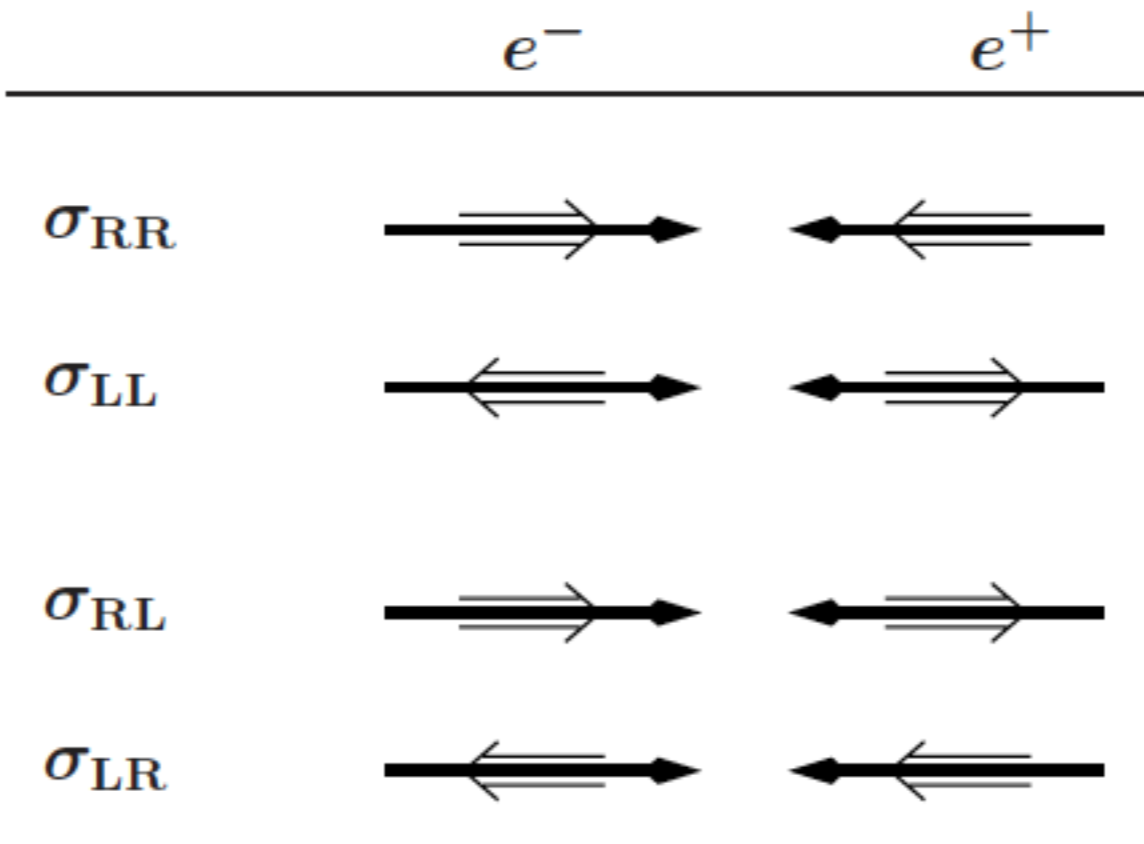}
     \caption{\label{fig:intro:configs}}
  \end{subfigure}
  \hspace{0.05\linewidth}
  \begin{subfigure}[h]{0.7\linewidth}
     \includegraphics[width=0.95\linewidth]{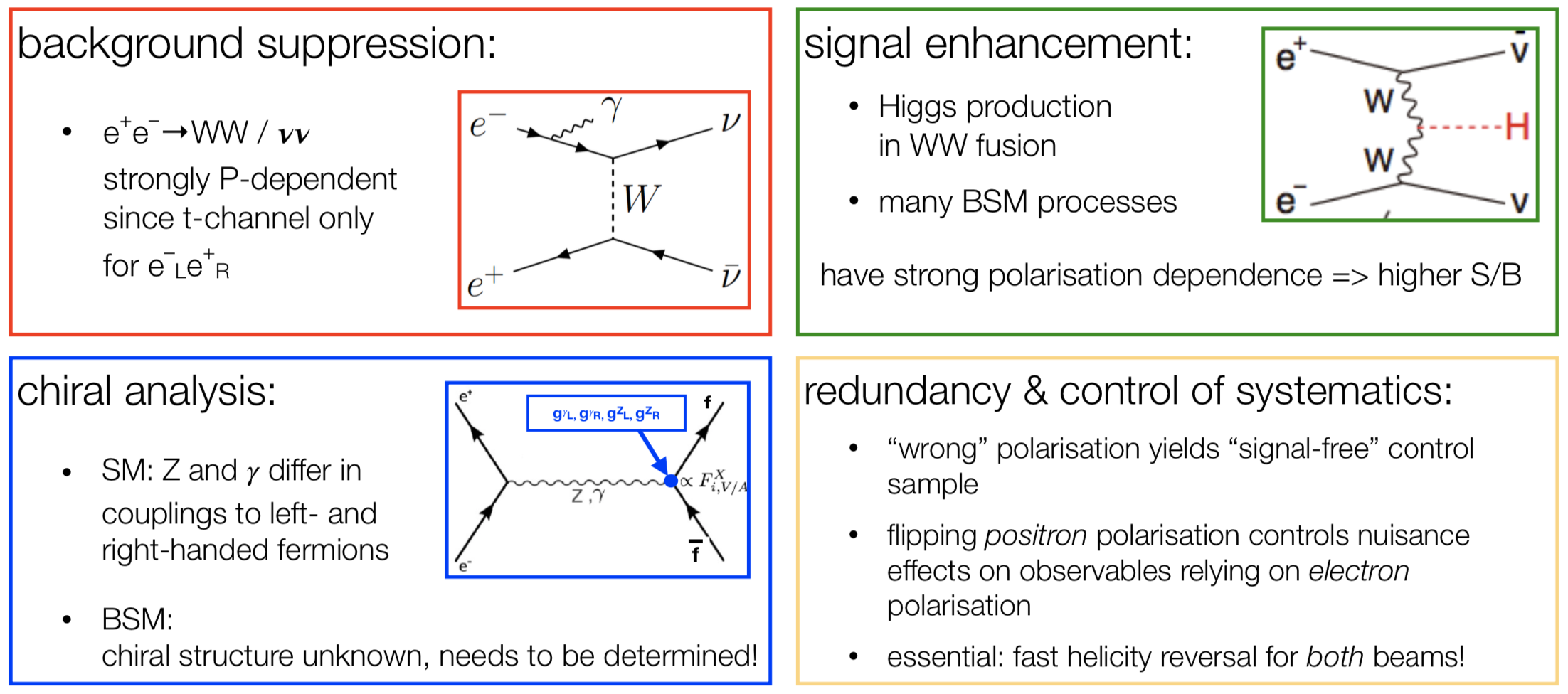} 
     \caption{\label{fig:intro:benefits}}
  \end{subfigure}
\caption{\label{fig:intro} (a) Spin configurations in collisions of longitudinally polarised electrons and positrons. (b) Overview of physics benefits of beam polarisation.}
\end{figure}

Longitudinally polarised beams are a special feature of Linear Colliders and are considered an integral part of the physics case, since, due to their $SU(2)_L \times U(1)$ gauge structure, electroweak interactions are highly sensitive to the chirality of fermions. The polarisation is defined as the number asymmetry between right- and left-handed particles in a bunch, $P=+80\%$ corresponding to $90\%$ right-handed and $10\%$ left-handed particles.  For CLIC, the electron beam, polarised to a degree of $80\%$, is planned to collide with unpolarised positrons, as it was the case at the SLC. For ILC also the positron beam is foreseen to be polarised, to a degree of $30\%$, upgradable to $60\%$. With both its beams polarised, the ILC can thus be considered as ``four colliders in one'', depending on which of the four configurations displayed in figure~\ref{fig:intro:configs} is enriched by the chosen polarisation signs. Figure~\ref{fig:intro:benefits} summarizes the four
main benefits of polarised beams for the physics program of a future $e^+e^-$ collider. The first three, namely the enhancement of signals, the suppression of backgrounds and the provision of additional observables for the analysis of the chiral properties of SM and BSM processes have been addressed often in the literature, see e.g.\ ref.~\cite{MoortgatPick:2005cw} for an in-depth review and ref.~\cite{Fujii:2018mli} specifically for ILC at 250\,GeV. The forth one, the control of systematic uncertainties, is less widely known, but equally important. It hinges upon the fact that for each process, the ``wrong'' polarisation configuration yields a control sample with very little signal content. In combination with the capability
of the accelerator to revert the sign of the polarisation on much faster time-scales than those of typical nuisance effects (drifts of calibrations, changes of alignment, etc.), data sets with different helicity configurations can be taken under otherwise identical experimental conditions. This allows for a nearly perfect cancellation of systematic effects between data sets with different polarisation signs. Similar reasoning can be applied for theoretical uncertainties, since the prediction of the effect of reversing the polarisation sign decouples from many sources of theoretical uncertainties. In other words, theory and detector systematics can be traded against polarisation systematics, which could arise
from the finite knowledge of the actual beam polarisation values. These, however, can be very well controlled, as will be described in Sec.~\ref{sec:polarimetry}. It should be noted that the reversal of the positron polarisation provides the redundacy required 
for controling nuisance effects on observables which depend on the electron polarisation.

\section{Role of Polarisation in the Physics Case of Future $e^+e^-$ Colliders}
The left-right asymmetry $A_{\mathrm{LR}}(ZH)$ of the cross section for Higgsstrahlung, $e^+e^-\to ZH$, plays a very important role for 
constraining physics beyond the SM via effective field theory-based 
interpretations of Higgs precision measurements. Two of the relevant diagrams at dimension-6, namely the SM diagram with a $Z$ exchange and
the (hypothetic) diagram with a photon exchange behave oppositely under
spin reversal of the incoming electron (positron): while the SM diagram flips its sign, the photon-exchange diagram keeps its sign. Thus by measuring $A_{\mathrm{LR}}(ZH)$, the degeneracy of the relevant EFT operators can be lifted~\cite{Barklow:2017suo}. Quantitatively this additional information corresponds to a gain of about $2.5$ in luminosity: In Higgs coupling determinations, e.g.\ $2$\,ab$^{-1}$ of polarised yield the same level of statistical precision as $5$\,ab$^{-1}$ of unpolarised data~\cite{Bambade:2019fyw}.

\begin{figure}[htb]  
  \begin{subfigure}[h]{0.49\linewidth}
     \includegraphics[width=0.95\linewidth]{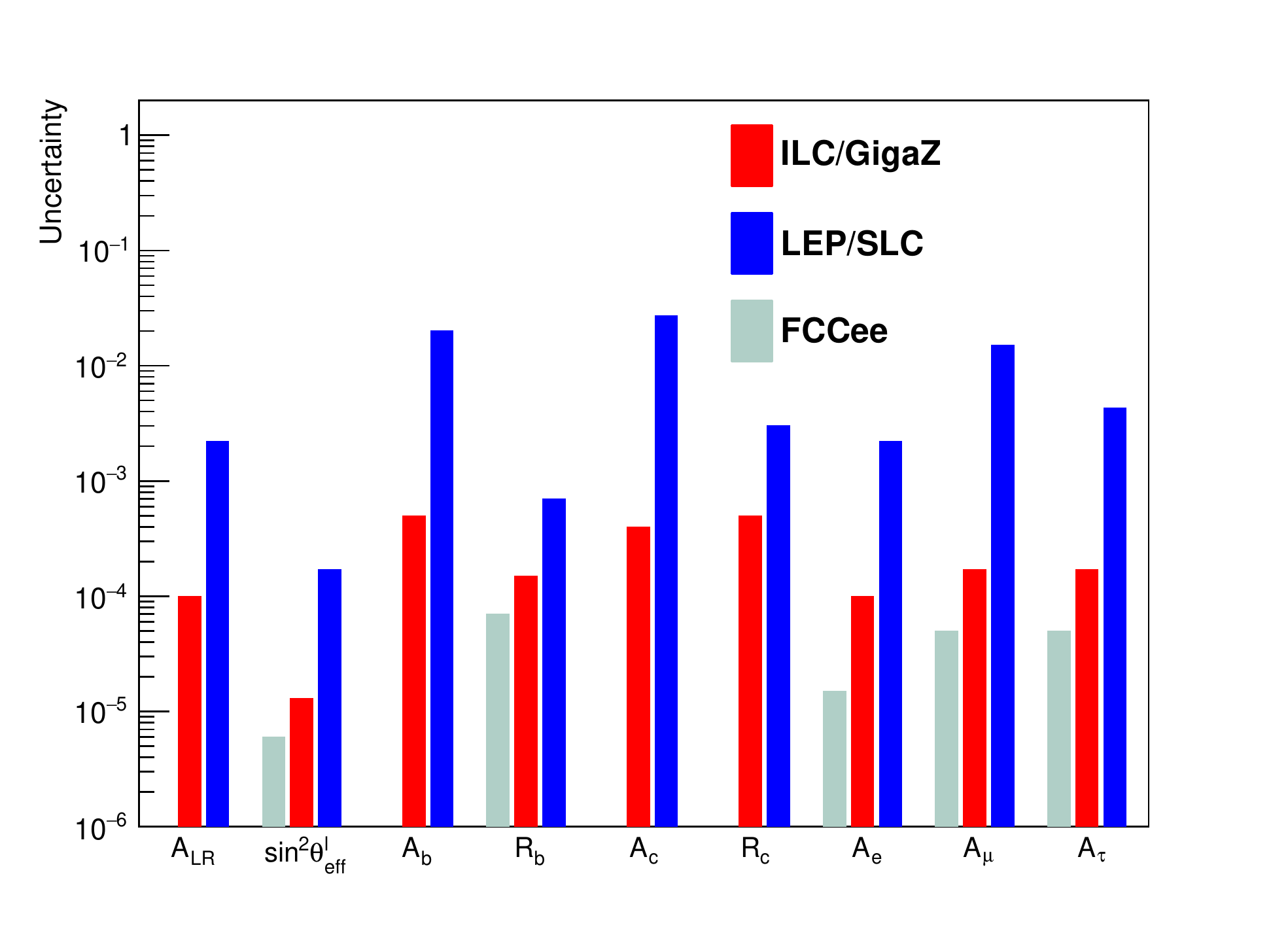}
     \caption{\label{fig:phys:Zpole}}
  \end{subfigure}
  \hspace{0.01\linewidth}
  \begin{subfigure}[h]{0.49\linewidth}
     \includegraphics[width=0.95\linewidth]{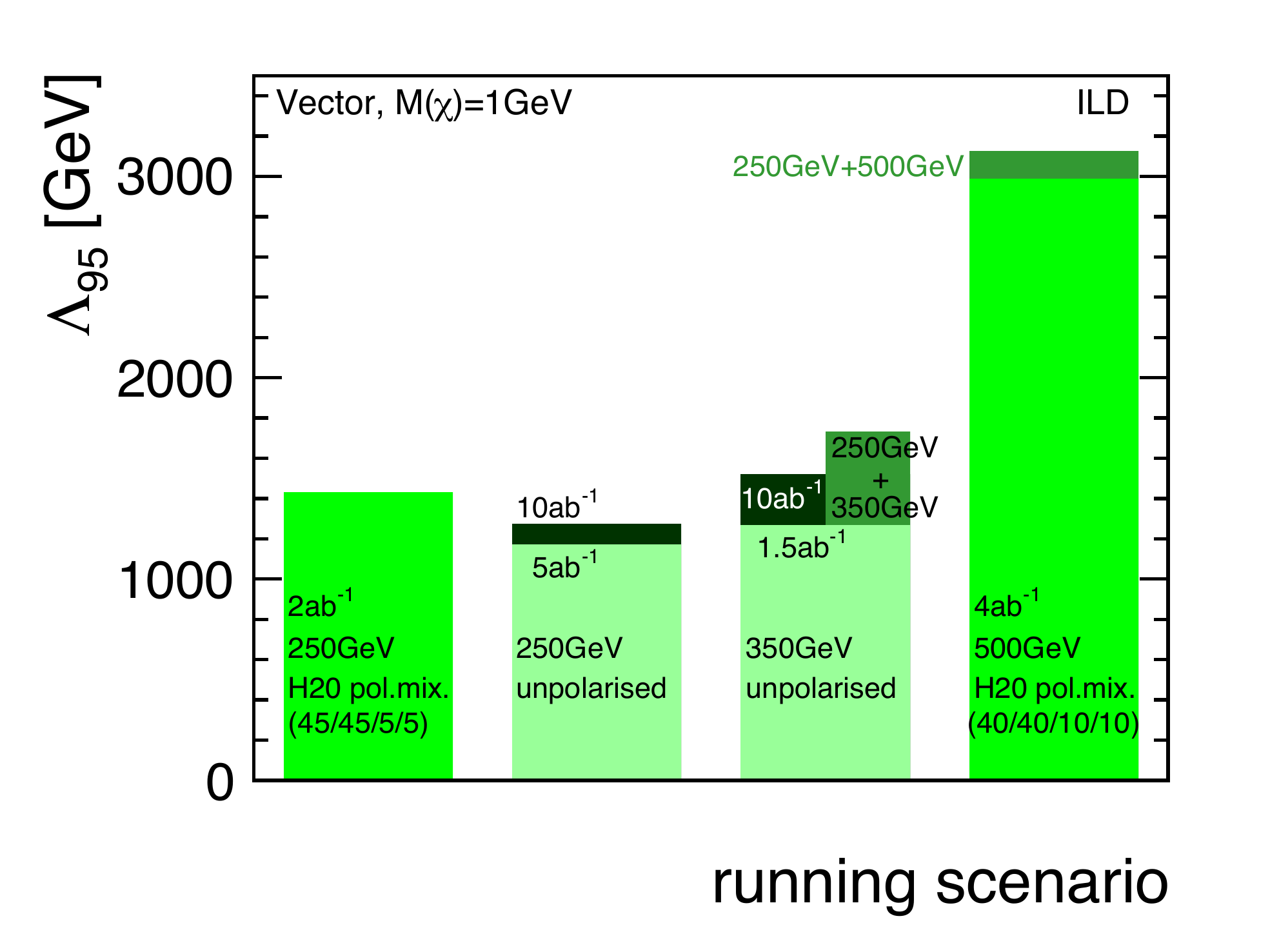} 
     \caption{\label{fig:phys:wimp}}
  \end{subfigure}
\caption{\label{fig:phys} (a) Expected precisions on various EWPOs 
from ILC's polarised GigaZ compared to current knowledge and FCCee's unpolarised TeraZ~\cite{Fujii:2019zll}. (b) Sensitivity reach in the energy scale for new physics $\Lambda$ for WIMP production mediated by a vector operator at the 95\% confidence level for various assumptions on center-of-mass energy, integrated luminosity and beam polarisation~\cite{Habermehl:2020njb}.}
\end{figure}

For bread-and-butter electroweak precision measurements like $e^+e^-\to f\bar{f}$, including $e^+e^-\to t\bar{t}$ at sufficiently high energies, beam polarisation both adds important information and can be traded against theory uncertainties. At the $Z$ pole, the asymmetry between the couplings of the right-handed and left-handed electron to the $Z$, $A_{\mathrm{e}}$, is
directly obtained from the left-right asymmetry of the total cross section, for which all visible $Z$ decay modes can be exploited. Compared to the extraction of $A_{\tau}$ from final-state polarisation of $\tau$-leptons from $Z\to\tau^+\tau^-$, this corresponds to an effective gain in luminosity of a factor of $100$, rendering typical precisions at ILC's GigaZ only a factor of about $3$ less precise than at FCCee's TeraZ, despite a factor $1000$ difference in assumed luminosity. As can be seen in figure~\ref{fig:phys:Zpole}, ILC would improve upon current knowledge by factors between 5 and 100, depending on the observable.
Above the $Z$ pole, e.g.\ in the Higgs factory mode at $250$\,GeV, beam polarisation is crucial in order to disentangle the contributions from
$Z$ and photon exchange, similar to the Higgs example above. Furthermore it turns out that the radiative corrections to the left-right-forward-backward asymmetry $A^f_{\mathrm{FB,LR}}$ are seven times smaller than those on the unpolarised forward-backward asymmetry $A^f_{\mathrm{FB}}$. Thus, polarised beams also lead to a reduction of
the impact of the residual theoretical uncertainties from higher orders~\cite{Fujii:2019zll}.

The power of combining data sets with different polarisation sign combinations for suppressing the impact of systematic uncertainties has been studied for instance in the search for WIMP production in the mono-photon channel. As shown in figure~\ref{fig:phys:wimp}, the sensitivity unpolarised data sets does hardly grow with additional luminosity. With the mix of polarisations signs foreseen at the ILC, already $2$\,ab$^{-1}$ outperform $10$\,ab$^{-1}$ unpolarised data. The other speciality of Linear Colliders, namely to reach higher energies, is obviously also very important for this kind of search, which probes energy scales up to $3$\,TeV at $\sqrt{s}=500$\,GeV, and up to $5$\,TeV at $\sqrt{s}=1$\,TeV.

\section{Polarimetry Concept at the ILC}
\label{sec:polarimetry}
In order to fully exploit the physics benefits of beam polarisation, the actual polarisation values need to be known very precisely, at the level of one permille. This is achieved for 
the ILC by combining the following items: a) Fast, time-resolved measurements of two Compton polarimeters per beam, $1.7$\,km up- and $140$\,m downstream of the $e^+e^-$ interaction point reach a systematic precision of $0.25\%$, with the statistical uncertainty becoming negligible on the timescale of a few seconds. The polarimeter design and the R\&D which demonstrates that the envisaged level of precision can be reached is summarized in ref.~\cite{Vormwald:2015hla}. b) The polarisation at the $e^+e^-$ interaction point can be obtained by interpolating the up- and downstream polarimeter measurents. Spin tracking calculations have demonstrated the ability to cross calibrate with a precision better than $0.1\%$~\cite{Beckmann:2014mka}, dominated by the relative alignment of orbit and spin vectors at the polarimeter locations. The depolarisation in collision will be monitored by comparing the up- and downstream measurements for colliding and non-colliding bunches. 
The time-resolved measurements will be averaged, weighted according to the instantaneous luminosity, to obtain the long-term averages relevant for the interpretation of collision data. c) The absolute scale of this long-term averages is in addition controlled
by the collision data themselves, e.g.\ by fitting the four average polarisation values
(electron beam and positron beam each with both signs) to cross sections and distributions which strongly depend on the polarisation -- either obtaining the polarisation values alone, or by treating them as nuisance parameters in a fit which extracts them along-side the actual physics observables. A prominent example often studied in the past is $W^+W^-$ production, where it is well established that the beam polarisations and charged triple gauge couplings can be extracted simultaneously without a net loss in precision~\cite{Marchesini:2011aka}. 

\begin{figure}[htb]  
  \begin{subfigure}[h]{0.49\linewidth}
     \includegraphics[width=0.95\linewidth]{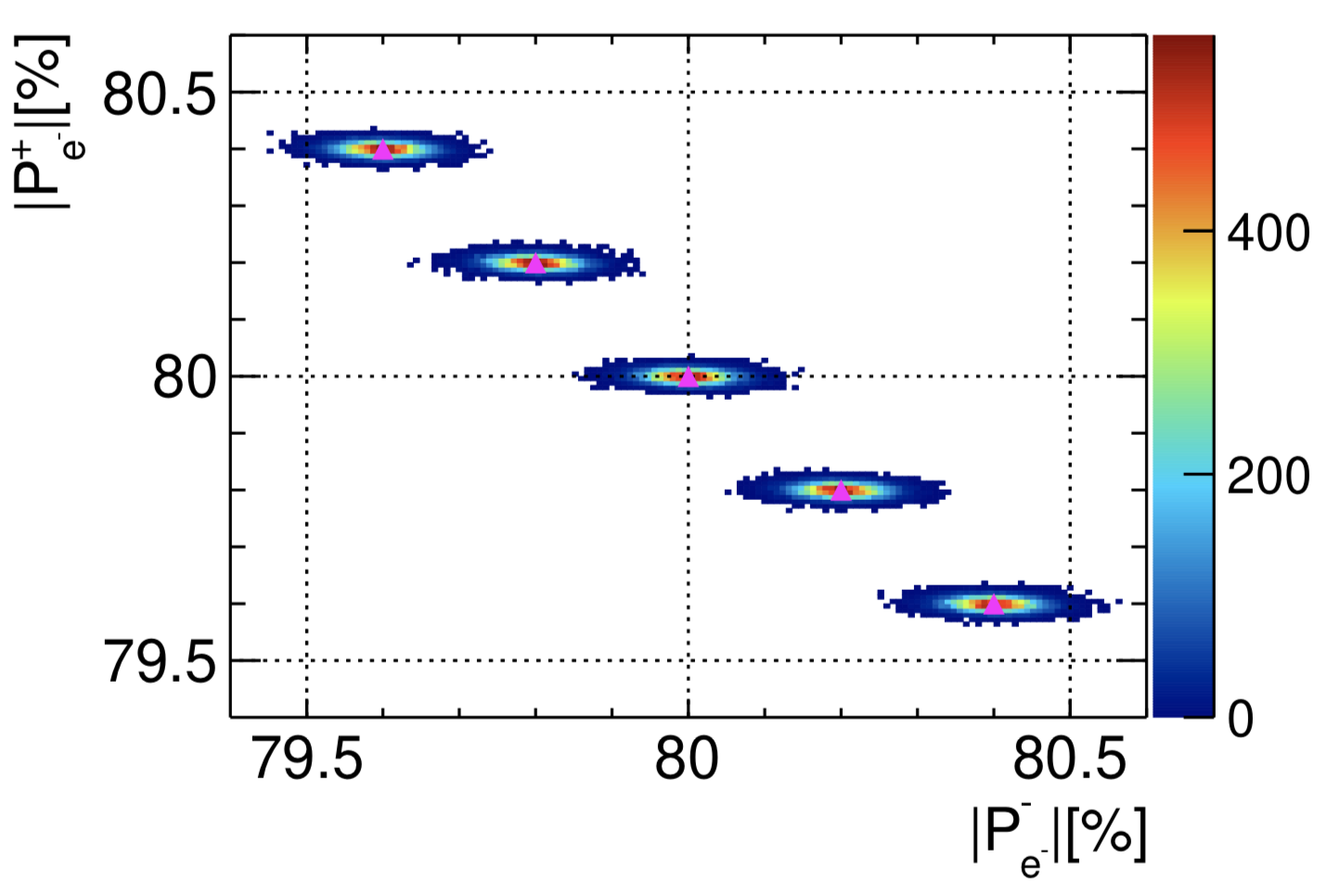}
     \caption{\label{fig:pol:nonequal}}
  \end{subfigure}
  \hspace{0.01\linewidth}
  \begin{subfigure}[h]{0.49\linewidth}
     \includegraphics[width=0.95\linewidth]{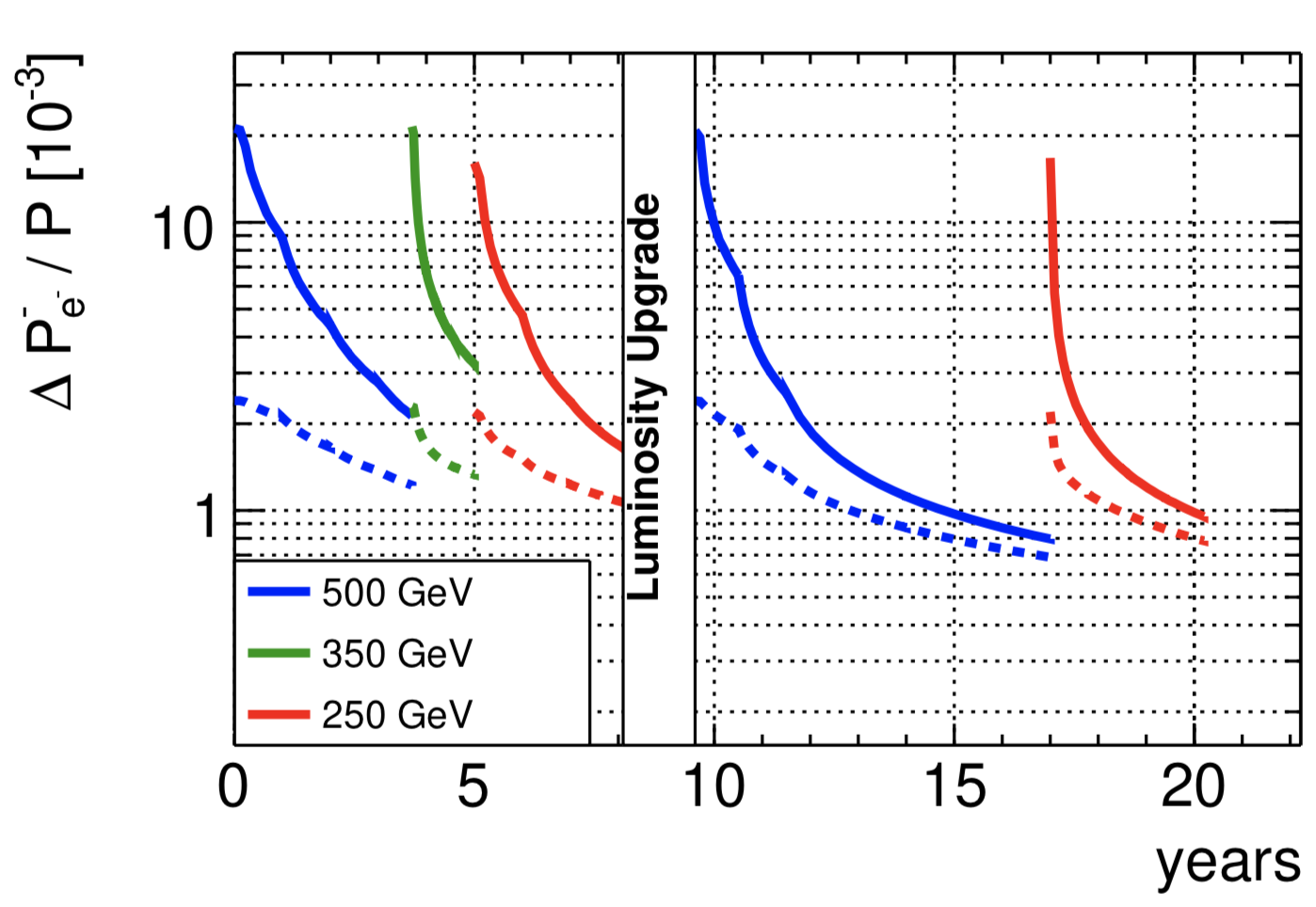} 
     \caption{\label{fig:pol:lumi}}
  \end{subfigure}
\caption{\label{fig:pol} (a) The absolute polarisation values do not need to be equal
when flipping the sign, since differences can be resolved with sufficient precision. Magenta triangles: input values; colour scale: number of fit outcomes on toy data sets.
(b) The time evolution of the precision of the average polarisations from collision data with operation time. The dashed lines show the improvement when including the polarimeter measurements. Here, the original ILC500 running scenario is assumed. In the staged scenario, only the order of runs is changed. Both from ref.~\cite{Karl:2019hes}.}
\end{figure}

More recently, the power of combining all relevant $e^+e^-\to f\bar{f}$ and $e^+e^-\to f\bar{f'}f''\bar{f'''}$ channels has been explored~\cite{Karl:2019hes}. Figure~\ref{fig:pol:nonequal} shows that in this approach even differences in the absolute polarisation values for positive and negative sign can be
reproduced with high precision, reproducing the input values indicated by the magenta triangles. Despite the power of the collision data, the additional information from the
polarimeters still plays an important role, as can be seen from the time evolution of the precision in figure~\ref{fig:pol:lumi}, where 
the full lines are obtained from collision data only, while the dashed lines include the
polarimeter measurements. 

With the combination of polarimeters, spin tracking and polarisation analysis of the collision data, the system provides sufficient redundancy to detect systematic biases in one of the methods, and to reach the goal of $0.1\%$ precision on the polarisation values~\cite{Karl:2019hes}.
It should be stressed that in particular for the redundancy aspect, the positron polarisation and the ability to reverse the polarisation inbetween bunch trains are essential.

\section{Conclusions}
Beam polarisation is an integral part of the physics case of future Linear Colliders like the ILC. Important examples comprise Higgs coupling measurements, top and electroweak physics at high energies, 
the $Z$ pole program as well as searches for production of new particles. All these require, however, the knowledge of the luminosity-weighted average polarisation values at the permille-level.
For the ILC, a detailed concept has been developed to reach this goal. It is based on time-resolved measurements 
of two Compton polarimeters per beam, spin-tracking across 
the 2\,km of beam delivery system and a global analysis of collision
data. With their combination, the polarisation is well under control, so that its physics benefits can be fully exploited.

\end{document}